\documentclass[a4paper,14pt]{article}
\usepackage{bm, color}
\usepackage{amsmath,amsfonts,amssymb,amsthm}
\usepackage{placeins}
\usepackage[title]{appendix}
\usepackage{systeme}
\usepackage{subfigure}
\usepackage{graphicx}
\usepackage{graphicx}
\usepackage{amsmath,amsfonts,amssymb,bm}
\usepackage{color}
\usepackage{ulem}
\usepackage{float}
\usepackage{caption}
\usepackage{multirow}
\usepackage[overload]{empheq}
\usepackage{cases}
\newcommand{\R}{\mathbb{R}}

\newtheorem{proposition}{Proposition}
\newtheorem{lemma}{Lemma}

\newtheorem{remark}{Remark}

\title{\bf Quantum MEP hydrodynamical model for  charge transport}
\author{V. D. Camiola,  V. Romano, G. Vitanza}
\date{}
\begin{document}
\maketitle
\begin{abstract}
A well known procedure to get quantum hydrodynamical models for charge transport is to resort to the Wigner equations and deduce the hierarchy  of the moment equations as in the semiclassical approach. 
If one truncates the moment hierarchy to a finite order, the resulting set of balance equations requires some closure assumption because the number of unknowns exceed the number of equations. 
In the classical and semiclassical kinetic theory a sound approach to get the desired closure relations is that based on the Maximum Entropy Principle (MEP) \cite{Jay1} (see\cite{CaMaRo_book}  for charge transport in semiconductors). 

In \cite{Ro} a quantum MEP hydrodynamical model has been devised for charge transport in the parabolic band approximation by introducing quantum correction based on the equilibrium Wigner function \cite{Wigner}. An extension to electron moving in pristine graphene has been obtained in \cite{LuRo}.

Here we present a quantum hydrodynamical model which is valid for a general energy band considering a closure of the moment system deduced by the Wigner equation resorting to a quantum version of  MEP.   Explicit formulas for quantum correction at order $\hbar^2$ are obtained with the aid of the Moyal calculus for silicon and graphene removing the limitation that the quantum corrections are based on the equilibrium Wigner function as in \cite{Ro,LuRo}. As an application, quantum correction to the mobilities are deduced.

%

\end{abstract}
\section{Introduction} 
 A way to study quantum transport is to resort to the Wigner function which leads to a description resembling the classical or semiclassical transport models.
 In fact, the mean values can be evaluated as  expectation values with respect to the Wigner function as it were a probability density.  The Wigner equation has been intensively investigated both from an analytical and numerical point of view  (the interested reader is referred to  \cite{MoSc, Mu,QuDo} and references therein) but it is almost exclusively assumed a quadratic dispersion relation for the
energy while  semiconductors or semimetal, e.g. graphene, obey different  dispersion
relations \cite{Ja,Ju,MaRo2}.  Quantum hydrodynamical models for charge transport in silicon
 have been devised  in \cite{Ro}  starting from the Wigner transport equation in the case of parabolic bands,
and in  \cite{LuRo2} the case of electrons moving in graphene has been tackled including quantum effects as second order corrections in the scaled Planck constant arising from the equilibrium Wigner function at the same temperature of  a thermal bath of phonons.


Quantum hydrodynamical models are obtained as moment equations of Wigner equation and as in the classical case, one gets  a system of balance equations which need some closure relations, that is one has to express the additional fields appearing in the moment equations in terms of a set of fundamental variables.  A sound way 
to accomplish this task is resorting to a quantum formulation of the maximum entropy principle \cite{Jay1} (hereafter QMEP). The quantum formulation of the maximum entropy principle was already given by Jaynes \cite{Jay2}. Recently,  a more formal theory has been developed in series of papers \cite{DeRi,DeMeRi} with several applications, for example for charge transport in semiconductors \cite{Ro,Barletti,BaCi,LuRo3}. The interested reader is also referred to \cite{Juen,CaMaRo_book}.  

Here we consider a general dispersion relation and formulate a hydrodynamic model for charge transport which is valid also beyond the parabolic band approximation. 
We apply QMEP to the moment equations deduced from the  Wigner one assuming as basic fields  those typical of a hydrodynamical model. that is  the density, energy density and momentum density  of  charge carriers.  Expanding up to second order in $\hbar$,  quantum corrections to the semiclassical case \cite{MaRo1} are deduced. Specific constitutive relations are explicitly obtained, with the aid of the Moyal calculus, for charge transport in silicon and graphene including the main electron-phonon scatterings. The results for graphene improve those obtained in \cite{LuRo3} in what concerns the  expression for the Lagrange multipliers of energy, no longer assumed to be that of a local thermal equilibrium with a bath of phonons. 
With a similar approach phonon transport has been tackled in \cite{CaRoVi}.

 The plan of the paper is as follows.  In section \ref{sec_Wigner} we write down the Wigner equations for charge transport. Section \ref{sec_moments} is dedicated  to deducing the moment equations whose closure relations are achieved by QMEP in section \ref{sec_QMEP}. Specific examples of hydrodynamical models for charge transport in bulk semiconductors and in graphene are presented in sections \ref{sec_hydro_sem} and \ref{sec_graph} respectively. In the last section quantum corrections to mobilities are obtained.
 
 
\section{Electron Wigner equation}\label{sec_Wigner}

The starting point of our derivation is the  one-particle Wigner function. 
For the system under consideration, given the density operator $\hat{\rho}$, i.e. a bounded non-negative operator with unit trace,
acting on $L^2(\mathbb{R}^d ,\mathbb{C})$, 
the associated Wigner function, $w = w({\bf x}, {\bf p}, t)$, evaluated at position ${\bf x}$, momentum  ${\bf p}, ({\bf x}, \, \, {\bf p}) \in \mathbb{R}^{2d}$, and time $t >0$, is the inverse Weyl quantization of $\hat{\rho}$,
\begin{equation}
w=Op^{-1}_{\hbar}(\hat{\rho}).
\end{equation}
We recall that the Weyl quantization of a phase-space function (a {\it symbol}) $a = a({\bf x}, {\bf p})$ is the (Hermitian) 
operator $Op_{\hbar}(a)$ formally defined by \cite{Hall}
\begin{equation}
Op_{\hbar}(a)\psi({\bf x}) = \frac{1}{(2\pi\hbar)^d}\int_{\mathbb{R}^{2d}}a\left(\frac{{\bf x}+{\bf y}}{2}, {\bf p}\right)\psi({\bf y})e^{i({\bf x}-{\bf y})\cdot {\bf p}/\hbar} d{\bf y} \, d {\bf p}
\end{equation}
for any $\psi \in L^2(\mathbb{R}^d ,\mathbb{C})$.
The inverse quantization of $\hat{\rho}$ can be written  as the {\it Wigner transform}
\begin{equation}
w({\bf x}, {\bf p}) = \frac{1}{\hbar^d}\int_{\mathbb{R}^d}\rho({\bf x}+\xi/2, {\bf x}-\xi/2)e^{i {\bf p}\cdot\xi/\hbar}d \xi,
\end{equation}
of the kernel  $\rho({\bf x}, {\bf y})$ of the density operator.

The dynamics of the time-dependent  Wigner function $w ({\bf x, p}, t)$ steams directly from the dynamics of the corresponding density operator $\hat{\rho}(t)$, i.e. from the
Von Neumann or quantum Liouville equation 
\begin{equation}\label{VonNeumannac}
i\hbar\partial_t \hat{\rho}(t) =[ \hat{H},\hat{\rho}(t)]:=\hat{H} \hat{\rho}(t)-\hat{\rho}(t) \hat{H},
\end{equation}
where $\hat{H}$ denotes the Hamiltonian operator  and $[\cdot, \cdot]$ the commutator. If $h = Op^{-1}_{\hbar}(\hat{H})$ is the symbol associated with $\hat{H}$, then, from Eq.s (\ref{VonNeumannac}), we obtain the {\it Wigner equation} 
\begin{equation}\label{Wignerac}
i\hbar\partial_t w ({\bf x, p}, t) = \{h, w({\bf x, p}, t)\}_{\#} := h \# w ({\bf x, p}, t)
 - w ({\bf x, p}, t) \# h. 
\end{equation}
With the symbol $\#$ we have denoted the  Moyal
(or {\it twisted}) product which translates the product of operators at the level of symbols according to
\begin{equation}\label{Moyal}
a \# b = Op^{-1}_{\hbar}(Op_{\hbar}(a)Op_{\hbar}(b)),
\end{equation}
for any pair of symbols $a$ and $b$. Here, we do not tackle the analytical issues which guarantee the existence of the 
previous relations but limit ourselves to remark that if two operators are in the Hilbert-Schmidt class, that is the trace there exists and it is not negative and bounded,  then the product is still Hilbert-Schmidt and the Moyal calculus is well defined. In the sequel, we will suppose that such conditions are valid. 

Let us consider the standard Hamiltonian symbol
\begin{equation}
h({\bf x},{\bf p})=\epsilon({\bf p})-q\Phi({\bf x})
\end{equation}
where $\epsilon({\bf p})$ is the energy band in terms of the crystal momentum $\bf{p} =\hbar \bf{k}$ and $\Phi({\bf x})$ the electrostatic  potential, which is assumed to be real. We do not assume for the moment a specific form of $\epsilon({\bf p})$. We have
\[
h\#w-w\#h = - i\hbar(S[\epsilon]w - q \Theta[\Phi]w)
\]
where
\begin{eqnarray}
& & i \hbar S[\epsilon]w = w\#\epsilon({\bf p})-\epsilon({\bf p})\#w,\\
& & i \hbar q \Theta[\Phi]w = q (w \#\Phi({\bf x})-\Phi({\bf x})\#w).
\end{eqnarray}

If we express the Moyal product as a power series
$$\epsilon({\bf p})\#w=\sum_{n=0}^{\infty}\hbar^n\epsilon({\bf p})\#_n w$$
we get (for the details see \cite{CaRoVi,BaCi})
\begin{eqnarray}
S[\epsilon]w =  \nabla_{\bf p}\epsilon({\bf p})\cdot \nabla_{\bf x}w - \dfrac{\hbar^2}{24}\partial_{\bf p}^3\epsilon({\bf p})\partial_{\bf x}^3w + O(\hbar^4),\\
\Theta[\Phi]w =   -\nabla_{\bf x}\Phi({\bf x})\cdot \nabla_{\bf p}w +\frac{\hbar^2}{24}\frac{\partial^3\Phi({\bf x})}{\partial x_i\partial x_j\partial x_k}\frac{\partial^3w}{\partial p_i\partial p_j\partial p_k}+O(\hbar^4).
\end{eqnarray}

Altogether, the Wigner equation reads 

\begin{equation}\label{Wigner1}
\frac{\partial}{\partial t} w({\bf x, p}, t) + S[\epsilon] w({\bf x, p}, t)-q\Theta[\Phi]w({\bf x, p}, t) = 0, 
\end{equation}


By adding a collision term, one has the 
the Wigner-Boltzmann equation 
\begin{equation}\label{Wigner2}
\frac{\partial}{\partial t} w({\bf x, p}, t)  + S[\epsilon] w({\bf x, p}, t) -q \Theta[\Phi] w({\bf x, p}, t) = C(w).
\end{equation}

We suppose that the expansion
\begin{equation}
w = w^{(0)}({\bf x, p}, t) + \hbar^2w^{(2)}({\bf x, p}, t) + O(\hbar^4)
\end{equation}
holds. Since formally as  $\hbar \to 0$ the semiclassical Boltzmann equation is recovered from the Wigner equation, $w^{(0)}({\bf x, p}, t)$ has to solve the first one while $w^{(2)}({\bf x, p}, t)$ is solution of the equation\footnote{Th Einstein convention on the repeated index is understood. }
\begin{eqnarray}
&&\frac{\partial}{\partial t}w^{(2)}({\bf x, p}, t) + \nabla_{\bf p}\epsilon({\bf p})\cdot \nabla_{\bf x} w^{(2)}({\bf x, p}, t) 
 - \dfrac{1}{24}
 \frac{\partial^3 \epsilon({\bf p})}{\partial p_i\partial p_j\partial p_k} \frac{\partial^3 w^{(0)}({\bf x, p}, t) }{\partial x_i\partial x_j\partial x_k}
 \\ &&+ q \left(\nabla_{\bf x}\Phi({\bf x})\cdot \nabla_{\bf p}w^{(2)}({\bf x, p}, t) - \frac{1}{24}\frac{\partial^3\Phi({\bf x})}{\partial x_i\partial x_j\partial x_k}\frac{\partial^3w^{(0)}({\bf x, p}, t)}{\partial p_i\partial p_j\partial p_k} \right) = C^{(2)}(w).
\end{eqnarray}
The explicit form of $C^{(2)}(w)$ will be specified later in the paper. 
\section{Moment equations} \label{sec_moments}
One of the most interesting properties of the Wigner function is that its moments have a
direct physical interpretation in terms of macroscopic fluid quantities, which makes Wigner
function an ideal tool for the derivation of quantum fluid equations. Analogously to previous works \cite{CaMaRo_book,LuRo3}, in this paper we shall
write equations involving the following moments to devise a quantum hydrodynamical model: 
\begin{itemize}
\item{the density 
\begin{equation}\label{vincolo1}
n({\bf x}, t) =y\int_{\mathbb{R}^d}w({\bf x, p}, t) d{\bf p},
\end{equation}}
\item{the momentum density
\begin{equation}\label{vincolo1}
{\bf J}({\bf x},t) = y \int_{\mathbb{R}^d} {\bf v} w({\bf x, p}, t) d{\bf p},
\end{equation}}
\item{the energy density
\begin{equation}\label{vincolo2}
W({\bf x},t) = y \int_{\mathbb{R}^d}\epsilon({\bf p})w({\bf x, p}, t) d{\bf p},
\end{equation}}
\end{itemize}
where $y=\dfrac{g_sg_v}{(2\pi\hbar)^d}$, $g_s$ and $g_v$ being spin and valley degeneracy, respectively.

The resulting moment system is obtained by taking the moments of the Wigner equation. 
Integrating  with respect to ${\bf p}$ the Wigner equation $(\ref{Wigner1})$ we have the following equation for the density \footnote{in the unipolar case the density production term is zero for the conservation of charge.}
$$
\frac{\partial}{\partial t}y\int_{\mathbb{R}^d}w({\bf x, p}, t) d{\bf p} +y\int_{\mathbb{R}^d}S[\epsilon]w({\bf x, p}, t) d{\bf p}-qy\int_{\mathbb{R}^d}\Theta[\Phi]w({\bf x, p}, t) d{\bf p} = 0, 
$$

Multiplying by $\epsilon({\bf p})$ and integrating with respect to ${\bf p}$ the  $(\ref{Wigner1})$ we obtain the equation for the energy density  $W$ 
$$
\frac{\partial}{\partial t}y\int_{\mathbb{R}^d}\epsilon({\bf p})w ({\bf x, p}, t) d{\bf p} +y\int_{\mathbb{R}^d}\epsilon({\bf p})S[\epsilon]w ({\bf x, p}, t) d{\bf p}-qy\int_{\mathbb{R}^d}\epsilon({\bf p})\Theta[\Phi] w({\bf x, p}, t)d{\bf p}=y\int_{\mathbb{R}^d}\epsilon({\bf p})Cd{\bf p}.
$$
while multiplying by ${\bf v}$ and integrating with respect to ${\bf p}$ the equation $(\ref{Wigner1})$ we obtain the equation for the  momentum density  ${\bf J}$  
$$
\frac{\partial}{\partial t}y\int_{\mathbb{R}^d}{\bf v}w({\bf x, p}, t) d{\bf p} +y\int_{\mathbb{R}^d}{\bf v}S[\epsilon]w({\bf x, p}, t) d{\bf p}-qy\int_{\mathbb{R}^d}{\bf v}\Theta[\Phi]w ({\bf x, p}, t) d{\bf p}=y\int_{\mathbb{R}^d}{\bf v}Cd{\bf p}.
$$

 Observe that up to terms of order in $\hbar^2$, $n=n^{(0)}+\hbar^2n^{(2)}$, where 
$$n^{(0)}=y\displaystyle{\int_{\mathbb{R}^d} w^{(0)}({\bf x}, {\bf p},t) d{\bf p}} \quad \mbox{and} \quad n^{(2)}=y\displaystyle{\int_{\mathbb{R}^d} w^{(2)}({\bf x}, {\bf p},t) d{\bf p}}.$$
 Similarly
$W=W^{(0)}+\hbar^2 W^{(2)}$ and ${\bf J} = {\bf J}^{(0)} + \hbar^2 {\bf J}^{(2)}$ with obvious meaning of the symbols.  


The moment equations split into zero and first order in $\hbar^2$ read

\begin{eqnarray}
& &\frac{\partial}{\partial t}n^{(0)} ({\bf x}, t)+\nabla_{\bf x}{\bf J}^{(0)}({\bf x}, t) = y\int_{\mathbb{R}^d}C^{(0)}d{\bf p}, \label{evolution01}\\
& &\frac{\partial}{\partial t} W^{(0)} + \nabla_{\bf x} {\bf S^{(0)}}-q\nabla_{\bf x}\Phi({\bf x})\cdot{\bf J^{(0)}}=y\int_{\mathbb{R}^d}\epsilon({\bf p})C^{(0)}d{\bf p},\\
& &\frac{\partial}{\partial t}{\bf J}^{(0)} + y\nabla_{\bf x}\int_{\mathbb{R}^d}{\bf v}\otimes{\bf v} \, w^{(0)}({\bf x, p}, t)  d{\bf p}
-y q\nabla_{\bf x}\Phi({\bf x})\int_{\mathbb{R}^d} w^{(0)}({\bf x, p}, t)\nabla_{\bf p}{\bf v} d{\bf p}=y\int_{\mathbb{R}^d}{\bf v}C^{(0)}d{\bf p}, \label{evolution03}
\end{eqnarray}
\begin{eqnarray}
& &\frac{\partial}{\partial t}n^{(2)}({\bf x}, t)+\nabla_{\bf x} {\bf J}^{(2)}({\bf x}, t)-y\frac{1}{24} \frac{\partial^3}{\partial x_i\partial x_j\partial x_k}\int_{\mathbb{R}^d}w^{(0)}({\bf x, p}, t)\frac{\partial^3 \epsilon({\bf p})}{\partial p_i\partial p_j\partial p_k}d{\bf p} = y\int_{\mathbb{R}^d}C^{(2)}d{\bf p},\\
& &\frac{\partial}{\partial t} W^{(2)} + \nabla_{\bf x} {\bf S}^{(2)}-q\nabla_{\bf x}\Phi({\bf x}) \cdot {\bf J}^{(2)} - \frac{y}{24}\frac{\partial^3}{\partial x_i\partial x_j\partial x_k}\int_{\mathbb{R}^d} \epsilon({\bf p})\frac{\partial^3 \epsilon({\bf p})}{\partial p_i\partial p_j\partial p_k} w^{(0)}({\bf x, p}, t) d{\bf p}
\nonumber\\
& &
+ q\frac{y}{24}\frac{\partial^3\Phi({\bf x})}{\partial x_i\partial x_j\partial x_k}\int_{\mathbb{R}^d}w^{(0)}({\bf x, p}, t)\frac{\partial^3 \epsilon({\bf p})}{\partial p_i\partial p_j\partial p_k}d{\bf p}=y\int_{\mathbb{R}^d}\epsilon({\bf p})C^{(2)}d{\bf p},\\
& &\frac{\partial}{\partial t} {\bf J}^{(2)} + y\left(\nabla_{\bf x}\int_{\mathbb{R}^d}{\bf v}\otimes{\bf v}w^{(2)}({\bf x, p}, t) d{\bf p}- \frac{1}{24}\frac{\partial^3}{\partial x_i\partial x_j\partial x_k}\int_{\mathbb{R}^d}{\bf v}\frac{\partial^3 \epsilon({\bf p})}{\partial p_i\partial p_j\partial p_k} w^{(0)}({\bf x, p}, t) d{\bf p}\right)\nonumber\\
& &-qy\nabla_{\bf x}\Phi({\bf x})\int_{\mathbb{R}^d} w^{(2)}({\bf x, p}, t) \nabla_{\bf p}{\bf v} d{\bf p}+q\frac{y}{24}\frac{\partial^3\Phi({\bf x})}{\partial x_i\partial x_j\partial x_k}\int_{\mathbb{R}^d}\frac{\partial^3 {\bf v}}{\partial p_i\partial p_j\partial p_k} w^{(0)}({\bf x, p}, t) d{\bf p} \nonumber\\
& &=y\int_{\mathbb{R}^d} {\bf v}C^{(2)}d{\bf p}, \label{eq_momento2}
\end{eqnarray}
where
$$
{\bf S}^{(0)}=y\displaystyle{ \int_{\mathbb{R}^d}\epsilon({\bf p}){\bf v}w^{(0)}({\bf x, p}, t) d{\bf p}} \quad \mbox{and} \quad {\bf S}^{(2)}=y \displaystyle{\int_{\mathbb{R}^d}\epsilon({\bf p}){\bf v}w^{(2)}({\bf x, p}, t) d{\bf p}}.
$$
are the zeroth and fist order terms in $\hbar^2$ of the energy flux. 

\section{QMEP for the closure relations}\label{sec_QMEP}
The evolution equations do not form a closed system of balance laws. We need to express the additional fields  as functions of $n$,  $W$ and ${\bf J}$. To this aim
the quantum version of the maximum entropy principle (QMEP) will be adopted. 

Since electrons are fermions, 
let us introduce the operator 
\begin{equation}\label{entropy}
s(\hat{\rho}) = -k_B[\hat{\rho} \ln \hat{\rho}+(1-\hat{\rho}) \ln(1- \hat{\rho})],
\end{equation}
which must be intended in the sense of the functional calculus. Here $k_B$ is the Boltzmann constant.
The entropy of the electrons reads 
$$S(\hat{\rho})=\mbox{Tr} \{s(\hat{\rho})\}$$
which can be viewed as a quantum Fermi-Dirac entropy.

According to MEP, we estimate $\hat{\rho}$ with $\hat{\rho}^{MEP}$ which is obtained by maximizing $S(\hat{\rho})$ under the constraints that some expectation values have to be preserved. In particular, in view of formulating hydrodynamical models we require that the following average values must be preserved
\begin{equation}\label{moments0}
 y \int_{\mathbb{R}^d}\bm{\psi}{({\bf p})}w^{MEP}({\bf x}, {\bf p},t) d{\bf p} = (n({\bf x}, t), W({\bf x}, t), {\bf J}({\bf x}, t)) := y \int_{\mathbb{R}^d}\bm{\psi}{({\bf p})}w({\bf x,p},t)d{\bf p},
\end{equation}
where
\begin{equation*}\label{weights}
\bm{\psi}{({\bf p})}=(1, \epsilon({\bf p}), \epsilon({\bf p}){\bf v})
\end{equation*}
is the vector of the weight functions
and $w^{MEP}$ is the Wigner function associated with $\hat{\rho}^{MEP}$ while $w$ is the Wigner function associated to $\hat{\rho}$. In the previous relations the time $t$ and position ${\bf x}$ must be considered as fixed.  In the moment conditions (\ref{moments0}), the first relation is a set of constraints while the second one is just a definition.  

The quantum formulation of MEP is given  in terms of expectation values. By taking into account that for a weight function $\psi ({\bf p})$ regular enough,
$$
\mbox{tr}\left\{ \hat{\rho} Op_{\hbar} (\psi ({\bf p}))\right\} (t) = y \int_{\mathbb{R}^d} \psi{({\bf p})}w({\bf x,p},t)d{\bf p} ,
$$
with the choice done above, the constrains read
\begin{eqnarray*} 
E_1 (t) = \mbox{tr}\left\{ \hat{\rho} Op_{\hbar} (1)\right\} (t), \quad
 {E}_2 (t) =  \mbox{tr}\left\{ \hat{\rho} Op_{\hbar} (\epsilon({\bf p}))\right\}(t),\quad
 {\bf E}_3 (t) =  \mbox{tr}\left\{ \hat{\rho} Op_{\hbar} ({\bf v})\right\}(t),
\end{eqnarray*}
and therefore, for fixed $t$,
\begin{eqnarray}
& &\hat{\rho}^{MEP} = \mbox{argument max} \, S(\hat{\rho}) \label{QMEP1}\\
& &\mbox{ under the constraints}\nonumber\\
& &\mbox{tr}\{ \hat{\rho}^{MEP} Op_{\hbar}  (1) \} = E_1 (t), \quad 
\mbox{tr}\{ \hat{\rho}^{MEP} Op_{\hbar} (\epsilon({\bf p})) \} =  {E}_2 (t), \label{QMEP2} \quad
\mbox{tr}\{ \hat{\rho}^{MEP} Op_{\hbar} ({\bf v}) \} =  {\bf E}_3 (t), \label{QMEP3}
\end{eqnarray}
in the space of the Hilbert-Schmidt operators on $L^2(\mathbb{R}^d ,\mathbb{C})$ which are positive, with trace one and such that the previous expectation values there exist.

If we introduce the vector of the Lagrange multipliers
\begin{equation}\label{multipliers}
\bm{\eta} = (\eta_{0}({\bf x}, t),{\eta}_{1}({\bf x}, t), {\bm\eta}_{2}({\bf x}, t)),
\end{equation}
the vector of the moments 
\begin{equation}\label{moments}
{\bf m[\rho](x,t)}:={\bf m}({\bf x}, t) = y\int_{\mathbb{R}^d}\bm{\psi} {({\bf p})}w ({\bf x,p},t)d{\bf p},
\end{equation}

and the vector of the  moments which must be considered as known
\begin{equation} \label{expectations}
{\bf M}({\bf x}, t):=\left(n({\bf x}, t), W({\bf x}, t) , {\bf J}({\bf x}, t)  \right),
\end{equation} 
the constrained optimization problem (\ref{QMEP1})-(\ref{QMEP2}) can be rephrased as a saddle-point problem for the Lagrangian
\begin{eqnarray}\label{objective}
{\cal L}(\hat{\rho},\bm{\eta}) &=& S(\hat{\rho}) - \int_{\mathbb{R}^d} \bm{\eta}  \cdot \left(
 {\bf m}({\bf x}, t) - {\bf M}({\bf x}, t) \right) \, d {\bf x} 
\nonumber\\
 & =& S(\hat{\rho}) - 
\mbox{tr}\left\{ \hat{\rho} Op_{\hbar} (\bm{\eta} \cdot(1,\epsilon({\bf p}), {\bf v}))\right\}  + \int_{\mathbb{R}^d} \bm{\eta} \cdot  {\bf M}({\bf x}, t)  \, d {\bf x} 
\end{eqnarray}
in the space of the admissible $\hat{\rho}$ and smooth function $\bm{\eta}$.

If the Lagrangian  ${\cal L}(\hat{\rho},\bm{\eta})$ is G\^ateaux-differentiable with respect to $\hat{\rho}$, the first order optimality condition
requires
$$
\delta {\cal L}(\hat{\rho}, \bm{\eta}) (\delta \hat{\rho}) = 0 
$$
for each Hilbert-Schmidt operator $\delta \hat{\rho}$  on $L^2(\mathbb{R}^d ,\mathbb{C})$  which is positive, with trace one and such that the previous expectation values there exist.

The existence of the first order G\^ateaux derivative is a consequence of the following Lemma 
\begin{lemma} \label{lemma1}
If $r(x)$ is a continuously differentiable increasing function on $\mathbb{R}^+$ then $\mbox{tr} \{ r(\hat{\rho})\}$ is G\^ateaux-differentiable in the class of the Hermitian Hilbert-Schmidt positive operators  on $L^2(\mathbb{R}^d ,\mathbb{C})$. The G\^ateaux derivative along 
$\delta \rho$ is given by
\begin{equation}
\delta \mbox{tr} \{ r(\hat{\rho})\} (\delta \hat{\rho}) = \mbox{tr} \left\{r'(\hat{\rho}) \delta \hat{\rho} \right\}.
\end{equation}
\end{lemma}
The extremality conditions for the unconstrained maximization problem (\ref{QMEP1})-(\ref{QMEP2}) are similar to that of the semiclassical case, as expressed by the following lemma.
\begin{lemma}
The first order optimality condition for the  maximization problem (\ref{QMEP1})-(\ref{QMEP2}) is equivalent to 
\begin{equation}
\hat{\rho}=(s')^{-1}(Op_{\hbar}(\bm{\eta} \cdot \bm{\psi}))
\end{equation}
where $(s')^{-1}$ is the inverse function of the first derivative of $s$.
\end{lemma}

$$
\delta {\cal L}(\hat{\rho}, \bm{\eta}) (\delta \hat{\rho}) = \mbox{tr} \left\{ \left(s'(\hat{\rho}) - Op_{\hbar}(\bm{\eta} \cdot \bm{\psi})\right) \delta \hat{\rho} \right\}
$$ 
$\forall \delta \hat{\rho}$ perturbation in the class of the Hermitian Hilbert-Schmidt positive operators  on $L^2(\mathbb{R}^d ,\mathbb{C})$. This implies
$$
s'(\hat{\rho}) = Op_{\hbar}(\bm{\eta} \cdot \bm{\psi}).
$$
\hfill $\Box$

Since the function $s(x)$ is concave, $s'(x)$ is invertible. Explicitly 
we have
\begin{equation*}
(s')^{-1}(z)=\frac{1}{e^{z/k_B}+1}
\end{equation*}
and  the operator solving the first order optimality condition reads
\begin{equation}
\hat{\rho}^* = (s')^{-1}(Op_{\hbar}(\bm{\eta} \cdot \bm{\psi}))=\frac{1}{e^{Op_{\hbar}(\bm{\eta}\cdot \bm{\psi})}+1}.
\end{equation}
Moreover, such an operator is a point of maximum for the Lagrangian.
\hfill $\Box$
\vskip 0.4cm

Now, to complete the program  we have to determine, among  the smooth functions, the Lagrange multipliers  $\bm{\eta}$ by solving the constraint
\begin{equation} 
\mbox{tr}\left\{ \hat{\rho} Op_{\hbar} (\bm{\eta} \cdot ( 1, \epsilon({\bf p}), {\bf v})\right\} - \int_{\mathbb{R}^d} \bm{\eta}  \cdot  {\bf M}({\bf x}, t)  \, d {\bf x} = 0. \label{vincolo_MEP}
\end{equation}
If such an equation admits a solution $\bm{\eta}^*$, the MEP density operator reads
\vskip 0.2cm
\begin{equation}
\hat{\rho}^{MEP} = 
         \frac{1}{\exp \left[Op_{\hbar}\left( \eta^*_{0}({\bf x},t)+{\eta}^*_{1}({\bf x},t) \epsilon({\bf p})+{\bm\eta}^*_{2}({\bf x},t) \cdot {\bf v}\right)  \right] +1},\label{solution}
\end{equation}
\vskip .2cm
\noindent where we have rescaled the Lagrange multipliers by the factor $1/k_B$. 

To determine conditions under which the equation (\ref{vincolo_MEP}) admits solutions is a very difficult task. Even in the semiclassical case
there are examples of sets of moments that cannot be moments of a MEP distribution.  

We will look for the solution up to first order in $\hbar^2$.
Once the MEP density function has been determined, the MEP Wigner function is given by 
$$w^{MEP}({\bf x}, {\bf p},t) = Op_{\hbar}^{-1}(\hat{\rho}^{MEP})$$
which can be used to get the necessary closure relations by evaluating the additional fields  with $w$ replaced by $w^{MEP}$.

We remark that the constraints (\ref{vincolo_MEP}) can be more conveniently expressed as 
$$
y  \int_{\mathbb{R}^{2d}} \bm{\eta}  \cdot  \bm{\psi}({\bf x}, t) w^{MEP}({\bf x}, {\bf p}, t) \, d {\bf p} \, d {\bf x} - \int_{\mathbb{R}^d} \bm{\eta}  \cdot  {\bf M}({\bf x}, t)  \, d {\bf x} = 0
$$
but we require, in analogy with the semiclassical case, the stronger conditions
$$
y\int_{\mathbb{R}^d}  \bm{\psi}({\bf x}, t) w^{MEP}({\bf x}, {\bf p}, t) \, d {\bf p}  = {\bf M}({\bf x}, t),
$$
where the Lagrange multipliers enter through  $w^{MEP}({\bf x}, {\bf p}, t)$.

\subsection{Determination of the Lagrange Multipliers}
We look formally for a solution in powers of $\hbar$ 
\begin{equation}
w^{MEP} = w^{MEP}_0 +\hbar w^{MEP}_1 +\hbar^2 w^{MEP}_2 + ... 
\end{equation}
firstly without taking into account the dependence of the Lagrange multipliers on $\hbar$.

Of course, on account of the properties of the Weyl quantization, $w^{MEP}_0$ is equal to the semiclassical counterpart 
\begin{equation}
w^{MEP}_0 =  \displaystyle{\frac{1}{1 + \exp (\xi) }}
\end{equation}
with 
$$
\xi=\eta_{0}({\bf x},t)+ {\eta}_{1}({\bf x},t)\epsilon({\bf p})+ {\bm\eta}_{2}({\bf x},t) \cdot {\bf v}
$$
where the Lagrange multipliers have been rescaled by the factor $1/k_B$.

On account of the properties of Moyal product \cite{BaCi}, $w^{MEP}_1 = 0$ and 
\begin{eqnarray}
w_2^{MEP} (\xi) =  \frac{e^{\xi}}{8(e^{\xi}+1)^3}\left[(1-e^{\xi})\left(\frac{\partial^2 {\xi}}{\partial x_i\partial x_j}\frac{\partial^2 {\xi}}{\partial p_i\partial p_j}-\frac{\partial^2 {\xi}}{\partial x_i\partial p_j}\frac{\partial^2 {\xi}}{\partial x_j\partial p_i}\right) 
\right.\nonumber \\
\left.+ \left(\frac{\partial^2 \xi}{\partial x_i\partial x_j}\frac{\partial \xi}{\partial p_i}\frac{\partial \xi}{\partial p_j}-2\frac{\partial^2 \xi}{\partial x_i\partial p_j}\frac{\partial\xi}{\partial p_i}\frac{\partial\xi}{\partial x_j}+\frac{\partial^2\xi}{\partial p_i\partial p_j}\frac{\partial\xi}{\partial x_i}\frac{\partial\xi}{\partial x_j}\right)\frac{(e^{2\xi}-4e^{\xi}+1)}{3(e^{\xi}+1)}\right].\qquad
\end{eqnarray}

Consistently, if we also expand the Lagrange multipliers as 
\begin{equation}
{\bm\eta} = {\bm\eta}^{(0)} + \hbar^2 {\bm\eta}^{(2)} + o(\hbar^2),
\end{equation}  
one has
\begin{eqnarray}
w_2^{MEP} (\xi^{(0)}, \xi^{(2)}) \simeq  - \frac{e^{\xi^{(0)}}}{(e^{\xi^{(0)}}+1)^2} \xi^{(2)} + \frac{e^{\xi^{(0)}}}{8(e^{\xi^{(0)}}+1)^3}\left[(1-e^{\xi^{(0)}})\left(\frac{\partial^2 {\xi^{(0)}}}{\partial x_i\partial x_j}\frac{\partial^2 {\xi^{(0)}}}{\partial p_i\partial p_j}-\frac{\partial^2 {\xi^{(0)}}}{\partial x_i\partial p_j}\frac{\partial^2 {\xi^{(0)}}}{\partial x_j\partial p_i}\right) 
\right.\nonumber \\
\left.+ \left(\frac{\partial^2 \xi^{(0)}}{\partial x_i\partial x_j}\frac{\partial \xi^{(0)}}{\partial p_i}\frac{\partial \xi^{(0)}}{\partial p_j}-2\frac{\partial^2 \xi^{(0)}}{\partial x_i\partial p_j}\frac{\partial\xi^{(0)}}{\partial p_i}\frac{\partial\xi^{(0)}}{\partial x_j}+\frac{\partial^2\xi^{(0)}}{\partial p_i\partial p_j}\frac{\partial\xi^{(0)}}{\partial x_i}\frac{\partial\xi^{(0)}}{\partial x_j}\right)\frac{(e^{2\xi^{(0)}}-4e^{\xi^{(0)}}+1)}{3(e^{\xi^{(0)}}+1)}\right],\qquad
\end{eqnarray}
where
\begin{eqnarray}
\xi^{(0)} &=& \eta_{0}^{(0)}({\bf x},t)+ {\eta}_{1}^{(0)}({\bf x},t)\epsilon({\bf p}) + {\bm\eta_2 }^{(0)}({\bf x},t) \cdot {\bf v}, 
\\ \xi^{(2)} &=& \eta_{0}^{(2)}({\bf x},t)+ {\eta}_{1}^{(2)}({\bf x},t)\epsilon({\bf p}) + {\bm\eta_2 }^{(2)}({\bf x},t) \cdot {\bf v}.
\end{eqnarray}

Let us suppose now a small anisotropy in the distribution by assuming that $\boldsymbol{\eta}_2$ is a quantity of order $\delta$ with $ |\delta | \ll1$. This can be also justified by observing that at equilibrium $\boldsymbol{\eta}_2 \approx {\bf 0}$. Therefore we formally write
\begin{align}
\xi^{(0)}=\eta_0^{(0)}({\bf x},t)+\eta_1^{(0)}({\bf x},t) \epsilon({\bf p})+\delta \boldsymbol{\eta}_2^{(0)}({\bf x},t) \cdot {\bf v} := \xi_0^{(0)}+
\delta \boldsymbol{\eta}_2^{(0)}({\bf x},t) \cdot {\bf v}, \nonumber \\
\xi^{(2)}=\eta_0^{(2)}({\bf x},t)+\eta_1^{(2)}({\bf x},t) \epsilon({\bf p})+\delta \boldsymbol{\eta}_2^{(2)}({\bf x},t) \cdot {\bf v} := \xi_0^{(2)}+
\delta \boldsymbol{\eta}_2^{(2)}({\bf x},t) \cdot {\bf v}.
\end{align}

Expanding also with respect to $\delta$, we get 
\begin{eqnarray}
w_0^{MEP} &=&\frac{1}{e^{\xi_0^{(0)}+\delta \boldsymbol{\eta}_2^{(0)}\cdot {\bf v}} +1} \simeq \frac{1}{e^{\xi_0^{(0)}}+1}-\frac{e^{\xi_0^{(0)}}}{(e^{\xi_0^{(0)}}+1)^2}
\delta \boldsymbol{\eta}_2^{(0)} \cdot {\bf v} \simeq w_{0,0}^{MEP} + \delta w_{0,1}^{MEP},\\
w_2^{MEP} &=& w_{2,0}^{MEP} + \delta w_{2,1}^{MEP},
\end{eqnarray}
where
\begin{eqnarray}
w_{2,0}^{MEP}=-\frac{e^{\xi_0^{(0)}}}{(e^{\xi_0^{(0)}}+1)^2} \xi_0^{(2)} +\frac{e^{\xi_0^{(0)}}}{8(e^{\xi_0^{(0)}}+1)^3} \left[ (1-e^{\xi_0^{(0)}})
\left( \frac{\partial^2\xi_0^{(0)}}{\partial x_i \partial x_j}
\frac{\partial^2\xi_0^{(0)}}{\partial p_i \partial p_j} - \frac{\partial^2\xi_0^{(0)}}{\partial x_i \partial p_j}  \frac{\partial^2\xi_0^{(0)}}{\partial x_j \partial p_i} \right) \right. \nonumber \\
 \left. + \left( \frac{\partial^2\xi_0^{(0)}}{\partial x_i \partial x_j}  \frac{\partial \xi_0^{(0)}}{\partial p_i} \frac{\partial \xi_0^{(0)}}{\partial p_j} 
-2 \frac{\partial^2\xi_0^{(0)}}{\partial x_i \partial p_j}  \frac{\partial \xi_0^{(0)}}{\partial p_i} \frac{\partial \xi_0^{(0)}}{\partial x_j}  
+ \frac{\partial^2\xi_0^{(0)}}{\partial p_i \partial p_j}  \frac{\partial \xi_0^{(0)}}{\partial x_i} \frac{\partial \xi_0^{(0)}}{\partial x_j}  \right) 
 \frac{e^{2\xi_0^{(0)}}-4e^{\xi_0^{(0)}}+1}{3(e^{\xi_0^{(0)}}+1)} \right],
\end{eqnarray}

\begin{eqnarray*}
&&w_{2,1}^{MEP}=-\frac{e^{\xi_0^{(0)}}}{(e^{\xi_0^{(0)}}+1)^2} \delta \boldsymbol{\eta}_2^{(2)} \cdot {\bf v} + \frac{e^{\xi_0^{(0)}}}{(e^{\xi_0^{(0)}}-1)^2} \xi_0^{(2)} \delta \boldsymbol{\eta}_2^{(0)} \cdot {\bf v} \nonumber \\
&+&\frac{e^{\xi_0^{(0)}}}{8(e^{\xi_0^{(0)}}+1)^3} \left\{ \left(1-e^{\xi_0^{(0)}} \right) \left( \frac{\partial^2 \xi_0^{(0)}}{\partial x_i \partial x_j}  \delta \boldsymbol{\eta}_2^{(0)} \cdot \frac{\partial^2 {\bf v}}{\partial p_i \partial p_j} + \frac{\partial^2 \xi_0^{(0)}}{\partial p_i \partial p_j} \delta \frac{\partial^2 \boldsymbol{\eta}_2^{(0)}}{\partial x_i \partial x_j} \cdot {\bf v}  \right. \right.\\
&-&\left. \left. \frac{\partial^2 \xi_0^{(0)}}{\partial x_j \partial p_i} \delta \frac{\partial \boldsymbol{\eta}_2^{(0)}}{\partial x_j} \cdot \frac{\partial {\bf v}}{\partial p_i} - \frac{\partial^2 \xi_0^{(0)}}{\partial x_j \partial p_i} \delta \frac{\partial \boldsymbol{\eta}_2^{(0)}}{\partial x_i} \cdot \frac{\partial {\bf v}}{\partial p_j}  \right)- e^{\xi_0^{(0)}} \delta \boldsymbol{\eta}_2^{(0)} \cdot {\bf v} \left( \frac{\partial^2 \xi_0^{(0)}}{\partial x_i \partial x_j} \frac{\partial^2 \xi_0^{(0)}}{\partial p_i \partial p_j}  \right. \right. \\
&-&\left. \left.  \frac{\partial^2 \xi_0^{(0)}}{\partial x_i \partial p_j} \frac{\partial^2 \xi_0^{(0)}}{\partial x_j \partial p_i}  \right) + \left[ 2\frac{\partial^2 \xi_0^{(0)}}{\partial x_i \partial x_j} \frac{\partial \xi_0^{(0)}}{\partial p_i} \delta \boldsymbol{\eta}_2^{(0)} \cdot \frac{\partial {\bf v}}{\partial p_j} + \frac{\partial \xi_0^{(0)}}{\partial p_i} \frac{\partial \xi_0^{(0)}}{\partial p_j} \delta \frac{\partial^2 \boldsymbol{\eta}_2^{(0)}}{\partial x_i \partial x_j} \cdot {\bf v} \right. \right. \\ 
&-&\left. \left.  2 \left( \frac{\partial^2 \xi_0^{(0)}}{\partial x_i \partial p_j} \frac{\partial \xi_0^{(0)}}{\partial p_i} \delta \frac{\partial \boldsymbol{\eta}_2^{(0)}}{\partial x_j} \cdot {\bf v} +  \frac{\partial^2 \xi_0^{(0)}}{\partial x_i \partial p_j} \frac{\partial \xi_0^{(0)}}{\partial x_j} \delta \boldsymbol{\eta}_2^{(0)} \cdot \frac{\partial {\bf v}}{\partial p_i} + \frac{\partial \xi_0^{(0)}}{\partial p_i} \frac{\partial \xi_0^{(0)}}{\partial x_j} \delta \frac{\partial \boldsymbol{\eta}_2^{(0)}}{\partial x_i} \cdot \frac{\partial {\bf v}}{\partial p_j} \right)  \right. \right. \\
&+& \left. \left. 2 \frac{\partial^2 \xi_0^{(0)}}{\partial p_i \partial p_j} \frac{\partial \xi_0^{(0)}}{\partial x_i} \delta \frac{\partial \boldsymbol{\eta}_2^{(0)}}{\partial x_j} \cdot {\bf v} +\frac{\partial \xi_0^{(0)}}{\partial x_i} \frac{\partial \xi_0^{(0)}}{\partial x_j} \delta \boldsymbol{\eta}_2^{(0)} \cdot \frac{\partial {\bf v}}{\partial p_i \partial p_j}  \right] \frac{e^{2\xi_0^{(0)}}-4 e^{\xi_0^{(0)}}+1 }{3(e^{\xi_0^{(0)}}+1)}  \right. \\ 
&+& \left. \frac{e^{\xi_0^{(0)}}(e^{2\xi_0^{(0)}}+2e^{\xi_0^{(0)}}-5)}{3(e^{\xi_0^{(0)}}+1)^2} \delta \boldsymbol{\eta}_2^{(0)} \cdot {\bf v} \left(  \frac{\partial^2 \xi_0^{(0)}}{\partial x_i \partial x_j} \frac{\partial \xi_0^{(0)}}{\partial p_i} \frac{\partial \xi_0^{(0)}}{\partial p_j} -2  \frac{\partial^2 \xi_0^{(0)}}{\partial x_i \partial p_j} \frac{\partial \xi_0^{(0)}}{\partial p_i} \frac{\partial \xi_0^{(0)}}{\partial x_j}  \right. \right. \\
&+&\left. \left.  \frac{\partial^2 \xi_0^{(0)}}{\partial p_i \partial p_j} \frac{\partial \xi_0^{(0)}}{\partial x_i} \frac{\partial \xi_0^{(0)}}{\partial x_j}  \right) \right\} \\
&+&\frac{e^{\xi_0^{(0)}}(1-2e^{\xi_0^{(0)}})}{8(e^{\xi_0^{(0)}}+1)^4} \delta \boldsymbol{\eta}_2^{(0)} \cdot {\bf v}  \left[ (1-e^{\xi_0^{(0)}})
\left( \frac{\partial^2\xi_0^{(0)}}{\partial x_i \partial x_j}
\frac{\partial^2\xi_0^{(0)}}{\partial p_i \partial p_j} - \frac{\partial^2\xi_0^{(0)}}{\partial x_i \partial p_j}  \frac{\partial^2\xi_0^{(0)}}{\partial x_j \partial p_i} \right) \right. \\
 &+&\left. \left( \frac{\partial^2\xi_0^{(0)}}{\partial x_i \partial x_j}  \frac{\partial \xi_0^{(0)}}{\partial p_i} \frac{\partial \xi_0^{(0)}}{\partial p_j} 
-2 \frac{\partial^2\xi_0^{(0)}}{\partial x_i \partial p_j}  \frac{\partial \xi_0^{(0)}}{\partial p_i} \frac{\partial \xi_0^{(0)}}{\partial x_j}  
+ \frac{\partial^2\xi_0^{(0)}}{\partial p_i \partial p_j}  \frac{\partial \xi_0^{(0)}}{\partial x_i} \frac{\partial \xi_0^{(0)}}{\partial x_j}  \right) 
 \frac{e^{2\xi_0^{(0)}}-4e^{\xi_0^{(0)}}+1}{3(e^{\xi_0^{(0)}}+1)} \right]
\end{eqnarray*}

The constraints  
\begin{eqnarray*}
 n({\bf x},t) &=& y\int_{\mathbb{R}^d}  w^{MEP}({\bf x},{\bf p},t) d{\bf p},\\[0.3cm]
W({\bf x},t) &=& y\int_{\mathbb{R}^d}\epsilon({\bf p}) w^{MEP}({\bf x},{\bf p},t) d{\bf p},\\[0.3cm]
{\bf J}({\bf x},t) &=& y\int_{\mathbb{R}^d} {\bf v}({\bf p}) \, \, w^{MEP}({\bf x},{\bf p},t) d{\bf p}.
\label{flux}
\end{eqnarray*}
can be split into two systems: one at zero order in $\hbar^2$
\begin{eqnarray}
 n^{(0)} ({\bf x},t) &=& y\int_{\mathbb{R}^d} w_{0}^{MEP} (\xi_0^{(0)}({\bf x},t), {\bf p}) d{\bf p}, \label{vincolo01} \\[0.3cm]
W^{(0)} ({\bf x},t) &=& y\int_{\mathbb{R}^d}\epsilon({\bf p}) w_{0}^{MEP} (\xi_0^{(0)}({\bf x},t), {\bf p}) d{\bf p},\\[0.3cm]
{\bf J}^{(0)} ({\bf x},t) &=& y\int_{\mathbb{R}^d} {\bf v}({\bf p}) \,\,w_{0}^{MEP} (\xi_0^{(0)}({\bf x},t), {\bf p}) d{\bf p},
\label{vincolo03}
\end{eqnarray}
and one at first order in $\hbar^2$ 
\begin{eqnarray}
 n^{(2)}  &=& y\int_{\mathbb{R}^d}  w_{2}^{MEP} (\xi^{(0)} ({\bf x},t), \xi^{(2)}({\bf x},t), {\bf p}) \,d{\bf p}, \label{vincolo21} \\[0.3cm]
W^{(2)}  &=& y\int_{\mathbb{R}^d} \epsilon({\bf p}) w_{2}^{MEP} (\xi^{(0)} ({\bf x},t), \xi^{(2)}({\bf x},t), {\bf p}) \, d{\bf p},\\[0.3cm]
{\bf J}^{(2)}  &=& y\int_{\mathbb{R}^d} {\bf v}  \,\, w_{2}^{MEP} (\xi^{(0)} ({\bf x},t), \xi^{(2)}({\bf x},t), {\bf p}) \, d{\bf p}.
\label{vincolo23}
\end{eqnarray}

The system (\ref{vincolo01})-(\ref{vincolo03}) is a set of nonlinear  algebraic equations; the system (\ref{vincolo21})-(\ref{vincolo23}) form a nonlinear system of PDEs for the Lagrange multipliers whose analytical solution seems very difficult to get. Indeed, the situation is even more cumbersome because in a numerical scheme the inversion of the constraints should be performed at each time step. 

It is possible to prove the following properties (see for example \cite{Ro}) 
\begin{proposition}
At zero order in $\hbar^2$ the map 
$ {\bm \eta} \to {\bf M}({\bm \eta})$  defined by the system  (\ref{vincolo01})-(\ref{vincolo03})
is (at least locally) invertible.  \label{nonlinear_inversion}
\end{proposition}
\begin{proposition}
The equations (\ref{evolution01})-(\ref{evolution03}) form a symmetric hyperbolic system of balance laws when the closure relation is that given by MEP.
\end{proposition}
In order to have an analytical guess about the solution of the system (\ref{vincolo01})-(\ref{vincolo03}) we adopt the same strategy used in \cite{Ro,LuRo3,CaRoVi} and expand the equations (\ref{vincolo01})-(\ref{vincolo03}) up to first order in $\delta$
\begin{eqnarray}
 n^{(0)} &\simeq& y\int_{\mathbb{R}^d}\frac{1}{e^{\xi_0^{(0)}}+1} d{\bf p},  \label{vincolo01bis} \\[0.3cm]
W^{(0)} &\simeq& y\int_{\mathbb{R}^d}\frac{ \epsilon({\bf p})}{e^{\xi_0^{(0)}} +1}d{\bf p},\\[0.3cm]
{\bf J}^{(0)} &\simeq& - y\int_{\mathbb{R}^d}  {\bf v} \displaystyle{\frac{ e^{\xi_0^{(0)}}  {\bm\eta}_{2}({\bf x},t) \cdot {\bf v}}{\left(e^{\xi_0^{(0)}} +1\right)^2}d{\bf p}} \label{vincolo03bis} 
\end{eqnarray}

Since we are requiring that for $\hbar = 0$ one recovers the semiclassical distribution, the approximation  is valid provided
\begin{eqnarray}
0 \le w_{0}^{MEP} \le 1
\iff
0 \le  \frac{1}{e^{\xi_0^{(0)}} + 1} - \frac{e^{\xi_0^{(0)}}  {\bm\eta}_{2}({\bf x},t) \cdot {\bf v}}{\left(e^{\xi_0^{(0)}} +1\right)^2} \le 1. \label{compatibility}
\end{eqnarray}

Now we can not apply directly proposition  \ref{nonlinear_inversion} because of the expansion. However, the Jacobian matrix 
$$
\frac{\partial \left(n^{(0)}, W^{(0)} \right)}{\partial \left(  \eta_{0}^{(0)},  \eta_{1}^{(0)} \right)} = 
\left(
\begin{array}{lr}
- y\displaystyle{\int_{\mathbb{R}^d}\frac{e^{\xi_0^{(0)}}}{\left(e^{\xi_0^{(0)}}+1\right)^2} d{\bf p}} & - y \displaystyle{\int_{\mathbb{R}^d}\frac{ e^{\xi_0^{(0)}} \epsilon({\bf p})}{\left(e^{\xi_0^{(0)}}+1\right)^2}} d{\bf p}\\
- y \displaystyle{\int_{\mathbb{R}^d}\frac{e^{\xi_0^{(0)}} \epsilon({\bf p})}{\left(e^{\xi_0^{(0)}}+1\right)^2}} d{\bf p} & - y \displaystyle{\int_{\mathbb{R}^d}\frac{e^{\xi_0^{(0)}} \epsilon^2({\bf p})}{\left(e^{\xi_0^{(0)}}+1\right)^2}} d{\bf p}
\end{array}
\right)
$$
is negative defined and therefore, by taking into account that the third equation is linear in ${\bm\eta}_{2}^{(0)}$,   we have what follows.
\begin{proposition}
The constraints  (\ref{vincolo01bis})-(\ref{vincolo03bis}) are at least locally invertible. \label{prop_invert}
\end{proposition}
 
Once the equations  (\ref{vincolo01})-(\ref{vincolo03}) or the approximated equations  (\ref{vincolo01bis})-(\ref{vincolo03bis}) are solved, the equations (\ref{vincolo21})-(\ref{vincolo23}) are a linear sytems for the second order correction in $\hbar$ to the Lagrange mltipliers. 

In the next section we will investigate some specific examples.

\section{Hydrodynamical model for charge transport in semiconductors} \label{sec_hydro_sem}
Let us consider a bulk (3d) semiconductor whose energy band in each valley can be approximated by the Kane dispersion relation
$$
\epsilon({\bf p}) \left( 1 + \alpha \epsilon({\bf p}) \right) = \frac{p^2}{2m^*} , \quad {\bf k} \in B,$$
where $B$ is the first Brillouin zone, expanded to all $\R^3$, $\alpha$ is the non parabolicity parameter and $m^*$ is the effective mass. The previous parameter depends on the specific material one are dealing with, e.g. silicon, GaAs, etc.. Sometimes the simple parabolic approximation ($\alpha=0$) is adopted. 

The group velocity is given by
$$
{\bf v} = \frac{1}{m^*\sqrt{1 + \dfrac{2 \alpha}{m^*} p^2}} {\bf p}.
$$ 
Note that it is limited; indeed
$$
|{\bf v}|\le v_{\infty}= \dfrac{1}{\sqrt{2 m^*\alpha}}.
$$
In the sequel we first study the first order terms and then the second order corrections in $\hbar^2$.
\begin{proposition} \label{propcond}
For the Kane dispersion relation a sufficient condition to satisfy (\ref{compatibility}) is  
\begin{equation}
| {\bm\eta}_{2} | \le \frac{ \left(e^{\xi_0} +1 \right)}{v_{\infty} e^{\xi_0^{(0)}}} \label{cond_compatibility}
\end{equation}
\end{proposition}
{\it Proof}. 
First we observe that
\begin{eqnarray*}
\frac{1}{e^{\xi_0^{(0)}}+1} - \frac{e^{\xi_0^{(0)}}  {\bm\eta}_{2}({\bf x},t) \cdot {\bf v}}{\left(e^{\xi_0^{(0)}} +1\right)^2} \ge 0 \iff 
{ \bm\eta}_{2}({\bf x},t) \cdot {\bf v} \le \frac{e^{\xi_0^{(0)}}+1}{e^{\xi_0^{(0)}}},
\end{eqnarray*}
which is satisfied if (\ref{cond_compatibility}) is true. 

On the other hand
\begin{eqnarray*}
 \frac{1}{e^{\xi_0^{(0)}} + 1} - \frac{e^{\xi_0^{(0)}}  {\bm\eta}_{2}({\bf x},t) \cdot {\bf v}}{\left(e^{\xi_0^{(0)}} +1\right)^2} \le 1 \iff 
 - { \bm\eta}_{2}({\bf x},t) \cdot {\bf v} \le e^{\xi_0^{(0)}} + 1,
\end{eqnarray*}
which surely holds if
\begin{eqnarray}
| {\bm\eta}_{2} | \le \frac{e^{\xi_0^{(0)}} +1}{v_{\infty}}.
\end{eqnarray} 
Since $\dfrac{ \left(e^{\xi_0^{(0)}} +1 \right)}{v_{\infty} e^{\xi_0^{(0)}}} \le \dfrac{e^{\xi_0^{(0)}} +1}{v_{\infty}}$, the proposition is proved.  
\hfill $\Box$

By taking into account the density of state for the Kane dispersion relation, one has
$$
d {\bf p} = p^2 d p \sin \theta \,  d \theta \, d \phi = (m^*)^{3/2} (1 + 2 \alpha \epsilon) \sqrt{2 \epsilon (1 + \alpha \epsilon)}  
d \epsilon \sin \theta \, d \theta \, d \phi, \quad \epsilon \in [0, +\infty[, \theta \in [0, \pi], \phi \in [0, 2 \pi].
$$
By using the relation
$$
\int_{S_2} {\bf n} \otimes {\bf n} \, d \, S_2 = \frac{4 \pi}{3} {\bf I}
$$
where $S_2$ is the unit sphere of $\R^3$ and ${\bf I}$ the identity tensor,  
and by taking into account that 
$$
\int_{S_2} \underbrace{{\bf n} \otimes {\bf n} \cdots \otimes {\bf n}}_{k \quad \mbox{times}} \, d \, S_2 = {\bf 0} \quad \mbox{if} \quad k \quad \mbox{odd},
$$
the constraints read
\begin{eqnarray}
n^{(0)} &=& 4 \pi y  \int_{0}^{+\infty}  (m^*)^{3/2} \frac{1 + 2 \alpha \epsilon}{\exp{(\eta_0^{(0)} + \eta_1^{(0)}  \epsilon)}+1} \sqrt{2 \epsilon (1 + \alpha \epsilon)}  
d \epsilon, \label{constraints_kane1}\\
W^{(0)} &=& 4 \pi y  \int_{0}^{+\infty}  (m^*)^{3/2} \frac{\epsilon (1 + 2 \alpha \epsilon)}{\exp{(\eta_0^{(0)} + \eta_1^{(0)}  \epsilon)}+1} \sqrt{2 \epsilon (1 + \alpha \epsilon)}  d \epsilon,\label{constraints_kane2}\\
{\bf J}^{(0)} &=& - \frac{8 \pi}{3} y \int_{0}^{+\infty} \frac{\exp{(\eta_0^{(0)} + \eta_1^{(0)}  \epsilon)} }{\left[\exp{(\eta_0^{(0)} + \eta_1^{(0)} \epsilon)}+1\right]^2}   
 \frac{\epsilon (1 +  \alpha \epsilon) (1 + 2 \alpha \epsilon)}{1+ 4 \alpha \epsilon (1 + \alpha \epsilon)} \sqrt{2 m^* \epsilon (1 + \alpha \epsilon)}  d \epsilon \, {\bm\eta}_{2}^{(0)}. \label{constraints_kane3}
\end{eqnarray}
The first two equations are a nonlinear system for the Lagrange multipliers $\eta_0$ and $\eta_1$ while the third equation gives ${\bm\eta}_{2}$ which results proportional to ${\bf J}^{(0)}$.

Thanks to proposition \ref{prop_invert}
equations (\ref{constraints_kane1}), (\ref{constraints_kane2}) are locally invertible. 
%
%


Once we have obtained the Lagrange multipliers (likely with a numerical procedure), we can evaluate the additional tensorial quantities appearing in eqs. (\ref{evolution01})-(\ref{evolution03}). Regarding the energy-flux, the pressure tensor and the  tensor gradient of the velocity, one has
\begin{eqnarray}
& &{\bf S}^{(0)} = y \int_{\mathbb{R}^3} \epsilon({\bf p})  {\bf v}({\bf p})  \, w_0^{MEP}({\bf x, p}, t)  d{\bf p} = \nonumber\\
& & - \frac{8 \pi}{3} y  \int_{0}^{+\infty} \frac{\exp{(\eta_0^{(0)} + \eta_1^{(0)}  \epsilon)} }{\left[\exp{(\eta_0^{(0)} + \eta_1^{(0)} \epsilon)}+1\right]^2} 
 \frac{\epsilon^2 (1 +  \alpha \epsilon) (1 + 2 \alpha \epsilon)}{1+ 4 \alpha \epsilon (1 + \alpha \epsilon)} \sqrt{2 m^* \epsilon (1 + \alpha \epsilon)}  d \epsilon \, {\bm\eta}_{2}^{(0)}. \\
& &\int_{\mathbb{R}^3}  {\bf v}\otimes{\bf v} \, w_0^{MEP}({\bf x, p}, t)  d{\bf p}
= \frac{8 \pi}{3} y \int_0^{+\infty} \frac{\epsilon (1 + \alpha \epsilon)  (1 + 2 \alpha \epsilon)}{1 + 4 \alpha \epsilon (1 + \alpha \epsilon)} 
 \frac{\sqrt{2 m^* \epsilon (1 + \alpha \epsilon)} }{\exp{(\eta_0^{(0)} + \eta_1^{(0)}  \epsilon)}+1} 
d \epsilon \,\, {\bf I} ,\\
& &\int_{\mathbb{R}^3} w_0^{MEP}({\bf x, p}, t)\nabla_{\bf p}{\bf v} d{\bf p} = \nonumber\\
& & 4 \pi \sqrt{m^*} y\int_0^{+\infty}\frac{(1 + 2 \alpha \epsilon) \sqrt{2 \epsilon (1 +  \alpha \epsilon) }}{(\exp{(\eta_0^{(0)} + \eta_1^{(0)}  \epsilon)}+1) \sqrt{1 + 4 \alpha \epsilon (1 + \alpha \epsilon)}} 
 \left[1 -  \frac{4 \alpha \epsilon (1 + \alpha \epsilon)}{3 (1 + 4 \alpha \epsilon (1 + \alpha \epsilon))} \right] d \epsilon \,\, {\bf I}.
\end{eqnarray}

In the parabolic band approximation the constraints can be written in terms of the Fermi integrals of order $k$
$$
{\cal F}_k (\eta) = \frac{1}{\Gamma (k+1)}\int_0^{+\infty} \frac{\chi^k}{1 + e^{\chi - \eta}} \, d \, \chi, 
$$
with $\Gamma(x)$ Euler gamma function, as
\begin{eqnarray}
n^{(0)} &=& 4 \pi y  \int_{0}^{+\infty}  (m^*)^{3/2} \frac{\sqrt{2 \epsilon}  }{\exp{(\eta_0^{(0)} + \eta_1^{(0)}  \epsilon)}+1} 
d \epsilon = \frac{y^*}{2 \left(\eta_1^{(0)} \right)^{3/2}} F_{1/2}(-\eta_0^{(0)}), \label{constraints_parab1}\\
W^{(0)} &=& 4 \pi y  \int_{0}^{+\infty}  (m^*)^{3/2} \frac{\epsilon \sqrt{2 \epsilon} }{\exp{(\eta_0^{(0)} + \eta_1^{(0)}  \epsilon)}+1}  d \epsilon =
\frac{3 y^* \sqrt{\pi}}{4 \left(\eta_1^{(0)} \right)^{5/2}} F_{3/2}(-\eta_0^{(0)}),\label{constraints_parab2}\\
{\bf J}^{(0)} &=& - \frac{8 \pi}{3 } y  \int_{0}^{+\infty} \frac{\exp{(\eta_0^{(0)} + \eta_1^{(0)}  \epsilon)} }{\left[\exp{(\eta_0^{(0)} + \eta_1^{(0)} \epsilon)}+1\right]^2} 
 \epsilon \sqrt{2 m^* \epsilon}  d \epsilon \, {\bm\eta}_{2}^{(0)}. \label{constraints_parab3}
\end{eqnarray}
where $y^* = 4 \pi y (m^*)^{3/2}  \sqrt{2}$.

Once the Lagrange multipliers have been determined at the zero order in $\hbar$, the constraints (\ref{vincolo21})-(\ref{vincolo23})  are  a linear system for 
$ {\bm\eta}^{(2)}$. 
In fact we can write
$$
w_{2}^{MEP}=-\frac{e^{\xi_0^{(0)}}}{(e^{\xi_0^{(0)}}+1)^2} \left(\xi_0^{(2)} + {\bm\eta}_{2}^{(2)} \cdot {\bf v}   \right) + \frac{e^{\xi_0^{(0)}}}{(e^{\xi_0^{(0)}}-1)^2} \xi_0^{(2)} \delta {\bm \eta}_2^{(0)} \cdot {\bf v} + \Psi ({\bm\eta}^{(0)}, \nabla {\bm\eta}^{(0)}).
$$
for a suitable scalar function $\Psi$.
Note that also the derivatives of $ {\bm\eta}^{(0)}$ enter the system. So, at variance with the semiclassical case, the constraints are nonlocal even by expanding in power of $\hbar$. 

The equations (\ref{vincolo21})-(\ref{vincolo23}) give
\begin{eqnarray}
 & &y\int_{\mathbb{R}^3} \frac{e^{\xi_0^{(0)}}}{(e^{\xi_0^{(0)}}+1)^2} \left(\eta_0^{(2)}+\eta_1^{(2)} \epsilon({\bf p})  \right)    d{\bf p}  =  - n^{(2)} + y \int_{\mathbb{R}^3} \Psi ({\bm\eta}^{(0)}, \nabla {\bm\eta}^{(0)})    d{\bf p}, \label{vincolo21bis} \\[0.3cm]
 & &y\int_{\mathbb{R}^3} \epsilon({\bf p}) \frac{e^{\xi_0^{(0)}}}{(e^{\xi_0^{(0)}}+1)^2} \left(\eta_0^{(2)}+\eta_1^{(2)} \epsilon({\bf p})  \right)  d{\bf p}   =  - W^{(2)}  + y \int_{\mathbb{R}^3} \epsilon({\bf p})  \Psi ({\bm\eta}^{(0)}, \nabla {\bm\eta}^{(0)})    d{\bf p}, \label{vincolo22bis} \\[0.3cm]
& &  \left(y\int_{\mathbb{R}^3} \frac{e^{\xi_0^{(0)}}}{(e^{\xi_0^{(0)}}+1)^2} {\bf v}  \otimes {\bf v} \,\, d{\bf p} \right)\,\, {\bm\eta}_{2}^{(2)}   = -  {\bf J}^{(2)}  + y\int_{\mathbb{R}^3} {\bf v}  \,\,  \Psi ({\bm\eta}^{(0)}, \nabla {\bm\eta}^{(0)})  d{\bf p}.
\label{vincolo23bis}
\end{eqnarray}
\begin{proposition}
The constraints relations (\ref{vincolo21bis})-(\ref{vincolo22bis}) are invertible. 
\end{proposition}
{\it Proof}. In fact, the matrix of the coefficients of the subsystem for $\eta_0^{(2)}$ and $\eta_1^{(2)}$
is the same of that appearing in Proposition \ref{prop_invert} and therefore invertible. Moreover the tensor
$$
\int_{\mathbb{R}^3} \frac{e^{\xi_0^{(0)}}}{(e^{\xi_0^{(0)}}+1)^2} {\bf v}  \otimes {\bf v} \,\, d{\bf p}
$$
is  positive defined. \hfill  $\Box$

Regarding the collisions, we require the conservation of charge in the unipolar case\footnote{in the bipolar case the presence of generation and recombination terms should be included.}. The general form of the scattering terms is rather cumbersome and very difficult to tackle. So, some simplification is adopted also in direct numerical integrations of the Wigner-Boltzmann equation.  An approach is to consider the collisions as semiclassical, that is the correction in $\hbar^2$ is neglected (an analysis about the validity of such an approximation can be found in \cite{QuDo}).

The main scattering mechanism in semiconductors is that between electrons and phonons. If the latter are considered as a thermal bath, and therefore obeying a  Bose-Einstein statistics with lattice temperature $T_L$,  the corresponding scattering has the expression
\begin{eqnarray}
C (w)  \simeq \frac{1}{(2 \pi)^3} \int_{\mathbb{R}^3} \left[ P ({\bf \tilde{p}},{\bf p}) w_{0}^{MEP} ({\bf \tilde{p}}) \left(1 -  w_{0}^{MEP}({\bf p}) \right) - P ({\bf p},{\bf \tilde{p}}) 
w_{0}^{MEP} ({\bf p})\left(1 -  w_{0}^{MEP} ({\bf \tilde{p}}) \right)\right]
\,\, d{\bf \tilde{p}}
\end{eqnarray}
where $P ({\bf \tilde{p}},{\bf p})$ is the transition probability per unit time to change the state of  momentum ${\bf \tilde{p}}$ into that of  momentum ${\bf p}$. The terms
$\left(1 -  w_{0}^{MEP}({\bf p}) \right)$ and $\left(1 -  w_{0}^{MEP}({\bf \tilde{p}}) \right)$ account for the Pauli exclusion principle. Note that $w_{0}^{MEP}$ is the semiclassical distribution, so the collision term is well defined. 

The general expression of the transition rate reads
$$
P ({\bf \tilde{p}},{\bf p}) = {\cal G} ({\bf \tilde{p}},{\bf p}) 
\left[ \left(N_B + 1 \right)\delta \left( \epsilon({\bf p})  - \epsilon({\bf \tilde{p}})  + \hbar \omega \right) 
+ N_B \delta \left( \epsilon({\bf p})  - \epsilon({\bf \tilde{p}})  - \hbar \omega \right)  \right]
$$
where ${\cal G} ({\bf \tilde{p}},{\bf p}) $ is the so-called {\cal overlap} factor, which enjoys the symmetry property ${\cal G} ({\bf \tilde{p}},{\bf p}) = {\cal G} ({\bf p},{\bf \tilde{p}})$, and $N_B$ is the Bose-Einstein distribution
$$
N_B = \frac{1}{\exp (\hbar \omega /k_B T_L) - 1},
$$
$k_B$ being the Boltzmann constant and $\delta$ the Dirac distribution. 
The first term represents an emission process of a quantum of energy $\hbar \omega$, the second one represents an absorption process of a quantum of energy $\hbar \omega$. 

In the specific case of silicon, one has ${\cal G} ({\bf \tilde{p}},{\bf p}) = \Lambda =$ constant\footnote{More in detail for the optical phonon scattering 
$\Lambda = Z_{if} \dfrac{\pi (D_T K)^2}{\rho \omega}$, where $Z_{if}$ is the degeneracy of the final valley, $\rho$ is the density of the material, $D_T K$ the optical coupling constant, $\omega$ the phonon angular frequency.}
and one gets the following expressions for the moments of the optical collision term $C_{op}$ in the Einstein approximation ($\hbar \omega =$ constant)  
\begin{eqnarray*}
&&C_n=0, \\
&&C_W= \frac{4 y \Lambda \hbar \omega m^{*3} N_B}{\pi}
\int_0^{+\infty} \left[ f_i(\epsilon, \epsilon+\hbar \omega)- f_i(\epsilon+\hbar \omega, \epsilon) e^{\hbar \omega / k_B T} \right]  (1+2\alpha \epsilon) (1+2 \alpha (\epsilon + \hbar \omega))  \\
&&\sqrt{\epsilon (\epsilon + \hbar \omega) (1+\alpha \epsilon) (1+\alpha (\epsilon + \hbar \omega))} \, d\epsilon, \\
&&C_{\bf V}=  \frac{y \Lambda m^{*3} N_B}{3 \pi^2} \int_0^{+\infty} (1+2 \alpha \epsilon)(1+2 \alpha (\epsilon + \hbar \omega)) \sqrt{\epsilon (\epsilon + \hbar \omega) (1+\alpha \epsilon) (1+\alpha (\epsilon + \hbar \omega))} \\
&& \left[ \left( E(\epsilon+ \hbar \omega) F_1(\epsilon, \epsilon+\hbar \omega)- E(\epsilon) F_2(\epsilon, \epsilon +\hbar \omega) \right) + e^{\hbar \omega / k_B T} \left( E(\epsilon) F_1(\epsilon+\hbar \omega, \epsilon) - E(\epsilon + \hbar \omega) F_2(\epsilon+\hbar \omega, \epsilon) \right) \right] \, d \epsilon \\
&&\delta \boldsymbol{\eta}_2^{(0)},
\end{eqnarray*}

with
\begin{eqnarray*}
f_i(\tilde{\epsilon},\epsilon)&=&\frac{1}{e^{\eta_0^{(0)}+\eta_1^{(0)} \tilde{\epsilon}}+1}\frac{e^{\eta_0^{(0)}+\eta_1^{(0)}\epsilon}}{e^{\eta_0^{(0)}+\eta_1^{(0)}\epsilon}+1}, \\
F_1(\tilde{\epsilon},\epsilon)&=& \frac{e^{\eta_0^{(0)}+\eta_1^{(0)}\epsilon}}{(e^{\eta_0^{(0)}+\eta_1^{(0)} \tilde{\epsilon}}+1)(e^{\eta_0^{(0)}+\eta_1^{(0)}\epsilon}+1)^2}, \\
F_2(\tilde{\epsilon},\epsilon)&=& \frac{e^{\eta_0^{(0)}+\eta_1^{(0)} \tilde{\epsilon}}}{e^{\eta_0^{(0)}+\eta_1^{(0)} \tilde{\epsilon}}+1} \left( \frac{1}{(e^{\eta_0^{(0)}+\eta_1^{(0)}\epsilon}+1)^2} - \frac{1}{e^{\eta_0^{(0)}+\eta_1^{(0)} \tilde{\epsilon}}} \right),  \\
E(\epsilon)&=&\sqrt{\frac{2 \epsilon (1+\alpha \epsilon)}{m[1+4 \alpha \epsilon (1+\alpha \epsilon)]}}
\end{eqnarray*}

In the limit $\hbar \omega \to 0$, one recovers the contribution to the production terms in the elastic approximation, often used for the acoustic phonons.

\begin{remark}
We remark that the use of MEP adopted here differs from the way it has been used in \cite{DeRi,DeMeRi,Barletti}. Indeed, in the quoted references the collision terms of the Wigner equations are written in a relaxation time form with a local equilibrium distribution which is estimated by MEP. This along with a Chapmann-Ensgok expansion allows to get macroscopic models like drift-diffusion, energy-transport or hydrodynamical ones. Here, we fix a set of fundamental variables and apply MEP to express the Wigner function in terms of the Lagrange multipliers. The resulting estimation is used to get closure relations both for the additional fluxes and production terms, without assuming for the latter a relaxation time form. 
\end{remark}
\section{Hydrodynamical model for charge transport in graphene} \label{sec_graph}
Now we apply the theory developed above to devise a hydrodynamical model for charge transport in graphene which is a 2D semimetal. 

Let us assume the following dispersion relation 
\begin{equation}
\varepsilon ({\bf p})=v_F \tilde{p}
\end{equation}
where $\tilde{p}=\sqrt{|{\bf p}|^2+c^2}$ with ${\bf p} \in  \mathbb{R}^2$, $c$ being a positive constant
that represents a possible half gap between the valence and the conduction band. Usually the gapless dispersion relation ($c = 0$) is considered but apparently a (even if small)  gap could be present from the general point of view \cite{CaNe}. 

The group velocity reads  
\begin{equation}
{\bf v}=\nabla_{{\bf p}} \varepsilon({\bf p})=\frac{v_F}{\tilde{p}} {\bf p}.
\end{equation}
The Proposition \ref{propcond} is still valid with $v_F$ instead of $v_{\infty}$.

Assuming also in this case small anisotropy, we get  the same expansions as for bulk (3d) semiconductors.  
The closure relations for the energy-flux,  pressure tensor and tensor gradient of the velocity at the zeroth order $\hbar^2$ in are given by

\begin{eqnarray*}
& &{\bf S}^{(0)} = -y \pi \int_{v_F c}^{+\infty} \frac{e^{\xi^{(0)}}}{(e^{\xi^{(0)}}+1)^2}(\epsilon^2 -v_F^2 c^2) d \epsilon \, \delta \boldsymbol{\eta}_0^{(0)}, \\
& &\int_{\mathbb{R}^2} {\bf v} \otimes {\bf v} w_0^{MEP}({\bf x},{\bf p},t) = \pi \int_{v_F c}^{+\infty} \frac{\epsilon^2-v_F^2 c^2}{\epsilon^2} \frac{1}{e^{\xi^{(0)}}+1} \, d\epsilon \, {\bf I}, \\
& &\int_{\mathbb{R}^2} w_0^{MEP}({\bf x},{\bf p},t) \nabla_{{\bf p}} {\bf v} = \pi \int_{v_F}^{+\infty} \left( 2 - \frac{\epsilon^2 -v_F^2 c^2}{\epsilon^2} \right) \frac{1}{e^{\xi^{(0)}}+1} d\epsilon \, {\bf I}.
\end{eqnarray*}
The second order corrections can be obtained with the same procedure for bulk semiconductors as well. 

Regarding the production terms, the main scattering mechanisms in graphene are those of electrons with acoustic, optical and $K$ phonons (see \cite{CaMaRo_book}). By using the MEP estimation of the Wigner function one can evaluate the closure relations for the production terms.

For the acoustic phonons one finds

\begin{eqnarray*}
&&C_n^{(ac)}=0, \\
&&C_W^{(ac)}=0,\\
&&C_{\bf V}^{(ac)}=-\frac{y A^{(ac)}}{4 \pi v_F^2} \int_{v_F c}^{+\infty} \epsilon \sqrt{\frac{\epsilon^2}{v_F^2}-c^2 } \frac{e^{\eta_0^{(0)} +\eta_1^{(0)} \epsilon}}{(e^{\eta_0^{(0)} +\eta_1^{(0)} \epsilon}+1)^2} d \epsilon  \boldsymbol{\eta}_2^{(0)},
\end{eqnarray*}
with $A^{(ac)}=\frac{2 \pi D^2_{ac}k_B T}{\sigma \hbar v^2_{ac}}$ where
$D^2_{ac}$ is the acoustic phonon coupling constant, $v_{ac}$ is the sound speed in graphene, $\sigma$ is the graphene areal density and $T_L$ is the graphene lattice temperature.

For the optical phonons we have 

\begin{eqnarray*}
&&C_n^{(opt)}=0,\\
&&C_W^{(opt)}=\frac{2y D^2_{\Gamma}}{\sigma \omega v_F^4}N_B \hbar \omega \int_{v_F C}^{+\infty} \epsilon (\epsilon +\hbar \omega) \left[ f_i(\epsilon,\epsilon+\hbar \omega) - e^{\frac{\hbar \omega}{k_B T}} f_i(\epsilon+\hbar \omega, \omega) \right] d \epsilon ,\\
&&C^{(opt)}_{\bf V}=\frac{y D^2_\Gamma N_B}{4 \sigma \omega \pi v_F^2} \\
&&\left\{ \int_{v_FC}^{+\infty} e^{\frac{\hbar \omega}{k_B T}}  \left[ F_1(\epsilon+\hbar \omega,\epsilon) G(\epsilon+\hbar \omega, \epsilon) - F_2(\epsilon+\hbar \omega,\epsilon) G(\epsilon, \epsilon+\hbar \omega)   \right] d \epsilon \right. \\
&&\left. + \int_{v_FC}^{+\infty} \left[ F_1(\epsilon, \epsilon+\hbar \omega ) G(\epsilon, \epsilon+\hbar \omega) -  F_2(\epsilon, \epsilon+\hbar \omega ) G(\epsilon+\hbar \omega, \epsilon) \right] d \epsilon \right\} \delta \boldsymbol{\eta}_2^{(0)},
\end{eqnarray*}
with
\begin{eqnarray*}
f_i(\tilde{\epsilon},\epsilon)&=&\frac{1}{e^{\eta_0^{(0)}+\eta_1^{(0)} \tilde{\epsilon}}+1}\frac{e^{\eta_0^{(0)}+\eta_1^{(0)}\epsilon}}{e^{\eta_0^{(0)}+\eta_1^{(0)}\epsilon}+1}, \\
F_1(\tilde{\epsilon},\epsilon)&=& \frac{e^{\eta_0^{(0)}+\eta_1^{(0)}\epsilon}}{(e^{\eta_0^{(0)}+\eta_1^{(0)} \tilde{\epsilon}}+1)(e^{\eta_0^{(0)}+\eta_1^{(0)}\epsilon}+1)^2}, \\
F_2(\tilde{\epsilon},\epsilon)&=& \frac{e^{\eta_0^{(0)}+\eta_1^{(0)} \tilde{\epsilon}}}{e^{\eta_0^{(0)}+\eta_1^{(0)} \tilde{\epsilon}}+1} \left( \frac{1}{(e^{\eta_0^{(0)}+\eta_1^{(0)}\epsilon}+1)^2} - \frac{1}{e^{\eta_0^{(0)}+\eta_1^{(0)} \tilde{\epsilon}}} \right),  \\
G(\tilde{\epsilon},\epsilon)&=& \frac{\tilde{\epsilon}}{\epsilon}(\epsilon^2 - v_F^2 C^2),
\end{eqnarray*}
where
$D^2_{\Gamma}$ is the optical phonon coupling constant, $N_B$ is the Bose-Einstein distribution
\begin{equation*}
N_B=\frac{1}{e^{\hbar \omega / k_B T_L}-1}
\end{equation*}
with $\hbar \omega$ the phonon energy (here assumed the same for LO and TO branches).

At last, the production terms given due to the optical K-phonons reads
\begin{eqnarray*}
&&C_n^{(K)}=0,\\
&&C_W^{(K)}=\frac{yD^2_{K}}{2 \sigma \omega v_F^4}N_B \hbar \omega \int_{v_F C}^{+\infty} \epsilon (\epsilon +\hbar \omega) \left[ f_i(\epsilon,\epsilon+\hbar \omega) - e^{\frac{\hbar \omega}{k_B T}} f_i(\epsilon+\hbar \omega, \omega) \right] d \epsilon, \\
&&C^{(K)}_{\bf V}=\frac{y D^2_K N_B}{4 \sigma \omega \pi v_F^2} \\
&&\left\{ \int_{v_FC}^{+\infty} e^{\frac{\hbar \omega}{k_B T}}  \left[ F_1(\epsilon+\hbar \omega,\epsilon) G(\epsilon+\hbar \omega, \epsilon) - F_2(\epsilon+\hbar \omega,\epsilon) G(\epsilon, \epsilon+\hbar \omega)   \right] d \epsilon \right. \\
&&\left. + \int_{v_FC}^{+\infty} \left[ F_1(\epsilon, \epsilon+\hbar \omega ) G(\epsilon, \epsilon+\hbar \omega) -  F_2(\epsilon, \epsilon+\hbar \omega ) G(\epsilon+\hbar \omega, \epsilon) \right] d \epsilon \right\} \delta \boldsymbol{\eta}_2^{(0)},
\end{eqnarray*}
where
$D^2_K$ is the K-phonon coupling constant.
\section{Quantum correction to mobility}
The model devised in the previous sections allow us to get corrections to the mobilities due to quantum effects. As well known, mobilities are crucial for the design of electron devices and, therefore, it is a crucial question to investigate the influence of quantum effects on them. 

In Sec.s \ref{sec_hydro_sem} and \ref{sec_graph}, the collisions have been considered as semiclassical, neglecting the terms in $\hbar^2$. To face with the complete second order correction of the collision operator in the Wigner equation is a formidable  task, so, in view of getting highlights on the mobilities, we adopt the following approach. The $\hbar^2$ correction of the production term of the  momentum balance equation (\ref{eq_momento2}) is modeled in a relaxation time form 
$$y\int_{\mathbb{R}^d}{\bf v} C  d{\bf p} = - \frac{ {\bf J}}{\tau}$$
where $\tau$ is a relaxation time
which can be obtained from (\ref{evolution03}) by imposing
\begin{equation*}
-\frac{{\bf J}^{(0)}}{\tau}=y\int_{\mathbb{R}^d} {\bf v} C^{(0)} d{\bf p}
\end{equation*}

Moreover, let us suppose to have a homogenous semiconductor undergoing a constant electric field, i.e. $\nabla_{\bf x}\Phi({\bf x})=$ constant, while the other spacial derivatives are zero. Under such an hypothesis, the  momentum balance equation up to order $\hbar^2$  reads
\begin{eqnarray*}
\frac{\partial}{\partial t} {\bf J} - q   y \int_{\mathbb{R}^d} \left[ w^{(0)}({\bf x, p}, t) + \hbar^2 w^{(2)}({\bf x, p}, t)\right] \nabla_{\bf p}{\bf v} d{\bf p} \, {\bf E} 
=  - \frac{ {\bf J}}{\tau}.
\end{eqnarray*}
In a long time scaling, one reaches a steady state and therefore
\begin{eqnarray} \label{steady_curr}
- \tau q y {\bf E} \int_{\mathbb{R}^d}  \left[ w^{(0)}({\bf x, p}, t) + \hbar^2 w^{(2)}({\bf x, p}, t)\right] \nabla_{\bf p}{\bf v} d{\bf p}
=   {\bf J}.
\end{eqnarray}
Recalling that the mobility $\boldsymbol{\mu}$, which is in general a tensor, is defined from the relation\footnote{We denote with $\boldsymbol{\mu} : {\bf E}$ the vector whose components are $v_k = \mu_{kh}E_h$.}
$$
 {\bf v} =  \boldsymbol{\mu} : {\bf E},
$$
and that
\begin{equation}
{\bf J} = n \boldsymbol{\mu} : {\bf E}.
\end{equation}
We expand the latter as 
\begin{eqnarray*}
{\bf J}^{(0)}+\hbar^2 {\bf J}^{(2)}&=&\left(  n^{(0)}+\hbar^2 n^{(2)} \right) \left( \boldsymbol{\mu}^{(0)}+ \hbar^2 \boldsymbol{\mu}^{(2)} \right) : {\bf E},\\
&=& n^{(0)} \boldsymbol{\mu}^{(0)} : {\bf E} + \hbar^2 \left( n^{(0)} \boldsymbol{\mu}^{(2)} + n^{(2)} \boldsymbol{\mu}^{(0)}  \right) : {\bf E} + o(\hbar^2).
\end{eqnarray*}
so
\begin{eqnarray*}
{\bf J}^{(0)}&=&n^{(0)} \boldsymbol{\mu}^{(0)},\\
{\bf J}^{(2)}&=&n^{(0)} \boldsymbol{\mu}^{(2)} + n^{(2)} \boldsymbol{\mu}^{(0)}.
\end{eqnarray*}

From (\ref{steady_curr}) one finds

\begin{eqnarray*}
\boldsymbol{\mu}^{(0)} &=& -\frac{\tau q y}{n^{(0)}}  \int_{\mathbb{R}^d} w_0^{MEP} \nabla_{\bf p} {\bf v} d {\bf p}, \\
\boldsymbol{\mu}^{(2)} &=&-\frac{n^{(2)}}{n^{(0)}}  \boldsymbol{\mu}^{(0)} - \frac{\tau q y}{n^{(0)}} \int_{\mathbb{R}^d} w_2^{MEP} \nabla_{\bf p} {\bf v} d {\bf p}.
\end{eqnarray*}

%

Specifically,  the contributions to mobility in the case of Kane dispersion relation reads

\begin{eqnarray*}
\boldsymbol{\mu}^{(0)}&=& - \frac{\tau q y}{n^{(0)}}  4 \pi \sqrt{m^*} y\int_0^{+\infty}\frac{(1 + 2 \alpha \epsilon) \sqrt{2 \epsilon (1 +  \alpha \epsilon) }}{(\exp{(\eta_0^{(0)} + \eta_1^{(0)}  \epsilon)}+1) \sqrt{1 + 4 \alpha \epsilon (1 + \alpha \epsilon)}} 
 \left[1 -  \frac{4 \alpha \epsilon (1 + \alpha \epsilon)}{3 (1 + 4 \alpha \epsilon (1 + \alpha \epsilon))} \right] d \epsilon \,\, {\bf I},\\
\boldsymbol{\mu}^{(2)} &=& -\frac{n^{(2)}}{n^{(0)}}  \boldsymbol{\mu}^{(0)} - \frac{\tau q y}{n^{(0)}}  4 \pi \left\{ \int_0^{\infty} \frac{e^{\xi_0^{(0)}}}{e^{\xi_0^{(0)}}+1} \xi_0^{(2)} \sqrt{\frac{2m^*\epsilon(1+\alpha \epsilon)}{1+4\alpha \epsilon (1+\alpha \epsilon)}}(1+2\alpha \epsilon)  d \epsilon  \right. \\
&&\left.-\frac{1}{3} \int_0^{\infty} \frac{e^{\xi_0^{(0)}}}{e^{\xi_0^{(0)}}+1} \xi_0^{(2)} \left[ 2m^*\epsilon (1+\alpha \epsilon) \right]^{5/2} m^*(1+2\alpha \epsilon) d \epsilon \right\} {\bf I},
\end{eqnarray*}
while in the case of graphene we get
\begin{eqnarray*}
\boldsymbol{\mu}^{(0)}&=&- \frac{\tau q y}{n^{(0)}} \pi \int_{cv_F}^{+\infty} \left( 2 - \frac{\epsilon^2 -v_F^2 c^2}{\epsilon^2} \right) \frac{1}{e^{\xi^{(0)}}+1} d\epsilon \, {\bf I},\\
\boldsymbol{\mu}^{(2)} &=&  -\frac{n^{(2)}}{n^{(0)}}  \boldsymbol{\mu}^{(0)} - \frac{\tau q y}{n^{(0)}} \frac{\pi}{v_F}\int_{cv_F}^{\infty} \frac{e^{\xi_0^{(0)}}}{e^{\xi_0^{(0)}}+1} \xi_0^{(2)} \left( 2-\frac{\epsilon^2 -c^2v_F^2}{\epsilon^2} \right) \sqrt{\epsilon^2- c^2v_F^2} \, d\epsilon {\bf I}.
\end{eqnarray*}

\section*{Conclusions and acknowledgements}
The Wigner equation for electrons moving in a $d$-dimensional crystal  has been written in the case of  a generic dispersion relation. Moment equations have been deduced and closed by QMEP taking into account the Fermi-Dirac statistics.  Explicit closure relations have been obtained for bulk semiconductors and for graphene under a suitable expansion both in $\hbar$ and in a parameter measuring the anisotropy of the Wigner functions. The devised models seem very useful for the simulation of charge transport in regime where quantum effects cannot be neglected.  

The authors acknowledge the support from INdAM (GNFM) and from MUR progetto PRIN
{\it Transport phonema in low dimensional structures: models, simulations and theoretical aspects} 
CUP E53D23005900006. V. D. Camiola acknowledges the financial support from  Universit\`a degli Studi di Catania, PIAno di inCEntivi per la RIcerca di Ateneo 2020/2022 - Linea di intervento 3 "Starting Grant".
%
%
%

\end{document}